\documentstyle[12pt,axodraw]{article}

\newcommand{\Li}[1]{\,{\rm Li}_{#1}}
\begin{document}

\thispagestyle{empty}
\onecolumn
\date{\today}
\vspace{-1.4cm}
\begin{flushleft}
{DESY 03-179 \\}
{BI-TP 2003/32  \\}
{SFB/CPP-03-57  \\}

\end{flushleft}
\vspace{1.5cm}

\begin{center}

{\LARGE {\bf
Analytical result for the two-loop QCD correction
to the decay $H \rightarrow 2\gamma$
        }
}
\vspace{1.5cm}

\vfill
{\large
 J. Fleischer$^{a}$,
 O.V. Tarasov$^{b,}$\footnote{On leave of absence from JINR,
141980 Dubna (Moscow Region), Russian Federation.}$^,\!\!$
\footnote{
Supported by the Deutsche Forschungs Gemeinschaft
under contracts FL 241/4-2 and SFB/TR9-03}
  and V.O.~ Tarasov$^{c}$%
}

\vspace{2cm}
$^a$~~Fakult\"at f\"ur Physik~~~~~\\
Universit\"at Bielefeld \\
D-33615 Bielefeld, Germany

\vspace{1cm}

$^b$~~Deutsches Elektronen-Synchrotron DESY~~~~~ \\
Platanenallee 6, D--15738 Zeuthen, Germany\\

\vspace{1cm}

$^c$~~Moscow Institute for Physics and Engineering ~~~~~ \\
Moscow , Russian Federation\\
\end{center}

\vfill

\begin{abstract}
An analytic formula for the two-loop QCD correction
of the decay $H \rightarrow 2\gamma$ is presented. 
To evaluate all master integrals a `Risch-like'
algorithm was exploited.
\end{abstract}

\vfill
\newpage

\setcounter{footnote}{0}

\section{Introduction}

   The introduction of the Higgs particle into the Standard Model 
allows to describe the masses of all other particles by means of 
spontaneous symmetry breaking. The `hidden' gauge symmetry also makes
the model renormalizable. In this sense the Higgs particle is the 
crucial element in the SM of electroweak interactions. The mass of 
the Higgs is still unknown and thus the search for the Higgs is a 
tremendous experimental task. At present one deduces all possible 
properties of the Higgs from perturbation theory even in higher 
orders to be prepared for the interpretation of coming experiments 
at LHC. Among these are decay modes of the Higgs, which have been 
investigated in two loop approximation by many authors (for  reviews 
see \cite{reviews}). Our contribution to this problem is in the 
present paper a complete analytic calculation of the decay 
$H \to 2 \gamma$ with the top quark as virtual loop.

\section{Results}

The partial decay width can be written in the form \cite{reviews}, 
\cite{SDGZ}
\begin{equation}
\Gamma[H \rightarrow \gamma \gamma]=
\left\vert \sum_{f}A_f(\tau_f)+A_W(\tau_W)\right\vert^2
\frac{M_H^3}{64 \pi},
\end{equation}
where $\tau_f=M^2_H/4m_f^2$, $\tau_W=M_H^2/4M_W$.
In the one loop order $A_W(\tau_W)$ is the dominating correction.
We will focus our attention on $A_t(\tau_t)$ and consider
QCD corrections of order $O(\alpha_s)$ to this amplitude.
It is convenient to write
\begin{equation}
A_t(\tau_t)=A^{(0)}_t+ \frac{\alpha_s}{\pi}A_t^{(1)}+ \ldots .
\end{equation}
The exact one-loop result is well-known \cite{ellis}, \cite{vainstein}
and reads
\begin{equation}
A^{(0)}_t= -\hat{A}_t
\left[ 1 - \frac{(1+\theta_t)^2}{4(1-\theta_t)^2}
 \ln^2\left(\frac{1}{\theta_t}\right) \right]
        \frac{6\theta_t}{(1-\theta_t)^2} ,
\end{equation}
where $\hat{A}_t=\frac{2\alpha}{3\pi v} N_C Q_t^2$,
$v=2^{-1/4}G_F^{-1/2}$, $N_C$ is the color factor, $Q_t$ is the electric
charge of the $t$- quark and

\begin{equation}
\theta_t=\frac{\sqrt{1-\frac{1}{\tau_t}}-1}
       {\sqrt{1-\frac{1}{\tau_t}}+1}.
\end{equation}

The two-loop diagrams contributing to $A^H_t(\tau_t)$ are shown
in Fig.1. There are six diagrams contributing to $H \gamma \gamma$
at the two-loop level (see Fig.1).

\begin{picture}(330,220)(0,0)
\Line(20,170)(90,200)
\Line(90,200)(90,140)
\Line(90,140)(20,170)
\Photon(90,200)(110,200){3}{3}
\Photon(90,140)(110,140){3}{3}
\Gluon(55,185)(55,155){2}{6}
\DashLine(5,170)(20,170){2}
\Text(107,190)[]{$q_2$}
\Text(107,150)[]{$q_1$}
\Line(150,170)(220,200)
\Line(220,200)(220,140)
\Line(220,140)(150,170)
\Photon(220,200)(240,200){3}{3}
\Photon(220,140)(240,140){3}{3}
\Gluon(185,185)(220,170){2}{8}
\DashLine(135,170)(150,170){2}
\Line(280,170)(350,200)
\Line(350,200)(350,140)
\Line(350,140)(280,170)
\Photon(350,200)(370,200){3}{3}
\Photon(350,140)(370,140){3}{3}
\Gluon(350,170)(315,155){2}{8}
\DashLine(265,170)(280,170){2}
\Line(20,70)(90,100)
\Line(90,100)(90,40)
\Line(90,40)(20,70)
\Photon(90,100)(110,100){3}{3}
\Photon(90,40)(110,40){3}{3}
\GlueArc(55,85)(15,20,200){2}{8}
\DashLine(5,70)(20,70){2}
\Line(150,70)(220,100)
\Line(220,100)(220,40)
\Line(220,40)(150,70)
\Photon(220,100)(240,100){3}{3}
\Photon(220,40)(240,40){3}{3}
\GlueArc(220,70)(15,90,270){2}{8}
\DashLine(135,70)(150,70){2}
\Line(280,70)(350,100)
\Line(350,100)(350,40)
\Line(350,40)(280,70)
\Photon(350,100)(370,100){3}{3}
\Photon(350,40)(370,40){3}{3}
\GlueArc(315,55)(15,-20,160){2}{8}
\DashLine(265,70)(280,70){2}
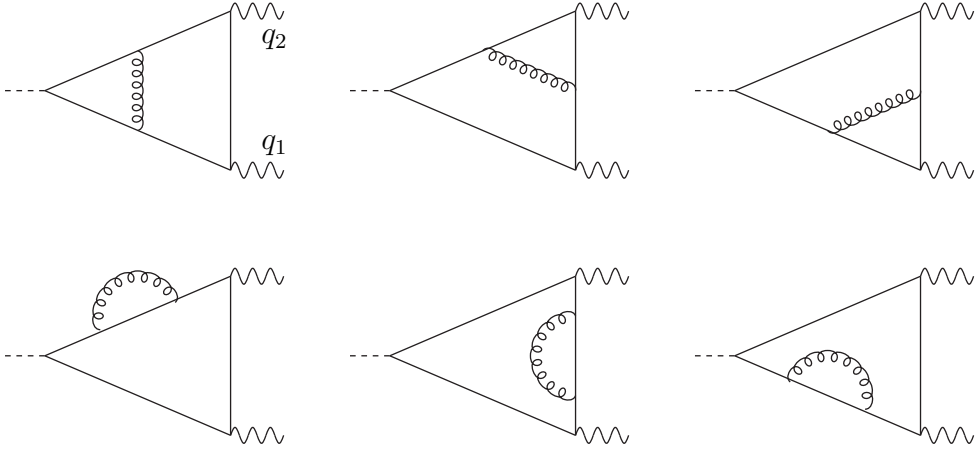
\Text(180,15)[]{Fig.1. Two-loop graphs contributing
   to the QCD  corrections to $H\gamma \gamma$
 decay.}

\end{picture}

The amplitude for the decay of the Higgs boson into two photons
with polarization vectors $\epsilon_\mu(q_1)$ and $\epsilon_\nu(q_2)$
has the following Lorentz structure:
\begin{eqnarray}
A_t^{\mu\nu}&=&\sum_{i} A_{t,i}^{\mu\nu}
          =\sum_{i} \left( a_{t,i}\, q_1q_2\, g^{\mu\nu}
                         + b_{t,i}\, q_1^{\nu}q_2^{\mu}
                         + c_{t,i}\, q_1^{\mu}q_2^{\nu} \right),
\label{amunu}
\end{eqnarray}
where $c_{t,i}$ has no contribution for on-shell photons.
In Eq.~(\ref{amunu}) the sum runs over all diagrams relevant for
the decay $H\to\gamma\gamma$.
Due to gauge invariance, we have $\sum_i a_{t,i} = -\sum_i b_{t,i}$.
It is easy to find projectors for $a_{t,i}$ and $b_{t,i}$:
\begin{eqnarray}
a_{t,i} &=& \frac{ A_{t,i}^{\mu\nu} }{(d-2)(q_1q_2)^2}
  \left(q_1q_2\, g_{\mu\nu} - q_{1\nu}q_{2\mu} - q_{1\mu}q_{2\nu}\right),\\
b_{t,i} &=& \frac{ A_{t,i}^{\mu\nu} }{(d-2)(q_1q_2)^2}
  \left(-q_1q_2\, g_{\mu\nu} + q_{1\nu}q_{2\mu} 
  + (d-1)q_{1\mu}q_{2\nu}\right).
\end{eqnarray}
We calculated both, $a_{t,i}$ and $b_{t,i}$,
in order to have an additional
check for the correctness of our result.

Tensor integrals and integrals with irreducible
numerators were represented in terms of scalar integrals
with shifted space-time dimension \cite{Tarasov:1996br}.

We introduced two auxiliary vectors $a_1$ and $a_2$
allowing to obtain tensors with $k_1$ and $k_2$ by
differentiating with respect to $a_{1,2}$,
using the differential operator
\begin{equation}
T_{\mu_1 \ldots \nu_1 \ldots } =\frac{1}{i^{n+k}} \frac{\partial^{n}}
                                    {\partial a_{1 \mu_1} \cdots
                                    \partial a_{1 \mu_n}}
\frac{\partial^{k}}
                                    {\partial a_{1 \nu_1} \cdots
                                    \partial a_{1 \nu_k}}
\exp ( i Q),
\label{Tdiff}
\end{equation}
where
\begin{eqnarray}
&&Q=-q_1a_1 (\alpha_1(\alpha_3+\alpha_4+\alpha_5+\alpha_6)
 +\alpha_3 \alpha_6 )
 -q_1 a_2 ( \alpha_3(\alpha_1+\alpha_2+\alpha_6+\alpha_7)
 +\alpha_2 \alpha_6 )
   \nonumber \\
&& \nonumber \\
&&~~-q_2 a_1(\alpha_2(\alpha_3+\alpha_4+\alpha_5+\alpha_6)
 +\alpha_4 \alpha_6)
 -q_2a_2( \alpha_4(\alpha_1+\alpha_2+\alpha_6+\alpha_7)
 +\alpha_2 \alpha_6)
   \nonumber \\
&& \nonumber \\
&&~~-\frac14a_1^2(\alpha_3+\alpha_4+\alpha_5+\alpha_6)
    -\frac12 a_1a_2 \alpha_6
    -\frac14a_2^2(\alpha_1+\alpha_2+\alpha_6+\alpha_7).
\end{eqnarray}

Applying the integration by parts method, e.g., to integrals with
selfenergy insertions in the fermion line `scratches' corresponding
fermion lines. For the remaining reduction of tensor integrals
we apply (\ref{Tdiff}) to the scalar integrals, setting $a_i=0$ after
differentiation. It is also convenient to consider integrals of a more
general type with the topology given in Fig.2. To deal with six-line
diagrams (Fig.1), the corresponding ${\alpha}_i$ are put to zero.

\begin{picture}(300,130)(0,0)
\Line(95,40)(245,40)
\Line(95,100)(245,100)

\Line(110,40)(110,100)
\Line(170,40)(170,100)
\Line(230,40)(230,100)
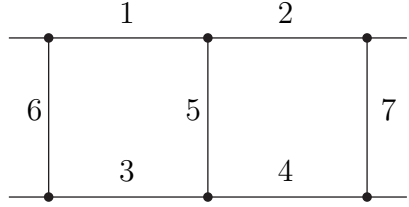
\Text(180,15)[]{Fig.2. Generic topology of the two-loop graphs}

\Text(140,110)[]{1}
\Text(140,50)[]{3}
\Text(200,110)[]{2}
\Text(200,50)[]{4}

\Text(105,73)[]{6}
\Text(165,73)[]{5}
\Text(239,73)[]{7}

\Vertex(110,40){1.8}
\Vertex(110,100){1.8}
\Vertex(170,40){1.8}
\Vertex(170,100){1.8}
\Vertex(230,40){1.8}
\Vertex(230,100){1.8}

\end{picture}

With the help of recurrence relations
one can reduce any integral with six lines to a combination
of integrals with at least one line contracted.
The two-loop correction in the on-shell renormalization
scheme reads

\begin{eqnarray}
&&A_t^{(1)}
 =-\hat{A}_t \left\{
        \frac{\theta_t(1+\theta_t)(1+\theta_t^2)}{(1-\theta_t)^5}
 \left[ \frac{1}{48} \ln^4\theta_t +\left( \frac72 \Li2(\theta_t)
        +2 \Li2(-\theta_t)\!+\!
          \zeta_2 \right) \ln^2 \theta_t\!+\!\frac92 \zeta_4
 \right. \right.
 \nonumber \\
&& \nonumber \\
&&~\left.-4\left(\Li3(\theta_t^2)-\zeta_3 \right) \ln \theta_t
   +36 \Li4(-\theta_t)+27 \Li4(\theta_t)
    \right ]+\frac{\theta_t(3\theta_t^3-7\theta_t^2
            +25\theta_t+3)}{12(1-\theta_t)^5} \ln^3\theta_t
 \nonumber \\
&& \nonumber \\
&&~~
    +\frac{\theta_t(5\theta_t^2-6\theta_t+5)}{2(1-\theta_t)^4}
                                \ln^2\theta_t \ln(1-\theta_t)
 +\frac{2\theta_t(3\theta_t^2-2\theta_t+3)}{(1-\theta_t)^4}
        \ln \theta_t \Li2(\theta_t)
   \nonumber \\
&& \nonumber \\
&&~~
 +\frac{\theta_t(1+\theta_t)^2}{(1-\theta_t)^4}
 \left[4 \ln \theta_t \Li2(-\theta_t)
 -\zeta_2 \ln \theta_t -8 \Li3(-\theta_t) \right]
   -\frac{\theta_t(7\theta_t^2-2\theta_t+7)}{(1-\theta_t)^4} \Li3(\theta_t)
     \nonumber \\
&& \nonumber \\
&&~~\left.
   +\frac{\theta_t(\theta_t^2-14\theta_t+1)}{(1-\theta_t)^4} \zeta_3
   +\frac{3\theta_t^2}{(1-\theta_t)^4} \ln^2 \theta_t
    -\frac{3\theta_t(1+\theta_t)}{(1-\theta_t)^3}
             \ln \theta_t - \frac{5\theta_t}{(1-\theta_t)^2}
                               \right\}.
\label{main} 			       
\end{eqnarray}
The expansion at $\tau_t=0$ agrees with the known 
result \cite{Dawson:1992cy}, \cite{Steinhauser:1996wy}:
\begin{eqnarray}
&&A^{(1)}_t=-\hat{A}_t \left(
  1-\frac{122}{135}\tau_t
 -\frac{8864}{14175}\tau_t^2
 -\frac{209186}{496125} \tau_t^3
 -\frac{696616}{2338875}\tau_t^4 \right. \nonumber \\
&&~~~\left.
 -\frac{54072928796}{245827456875}\tau_t^5
 -\frac{21536780128}{127830277575}\tau_t^6 - \ldots\right) .
\end{eqnarray}
and the asymptotic behavior near threshold $\tau_t \simeq 1$ is:
\begin{eqnarray}
&&A_t^{(1)}=-\hat{A}_t \left[
 \frac54-\frac{3}{16}\pi^2+\frac12 \pi^2 \ln 2
 -\frac74 \zeta_3 \right.
 \nonumber \\
&&~~+\left(2-\frac98 \pi^2 + \frac74 \pi^2 \ln 2
    + \frac{\pi^2}{4} \ln (1- \tau_t)
     \right) (1-\tau_t) \nonumber \\
&&~~~~~~~~~~~~~~\left.
     +{\it O}  ((1-\tau_t)^{3/2} \ln(1-\tau_t)) \right].
\end{eqnarray}

At $m_H^2> 4m_t^2$  the imaginary part of (\ref{main}) is in agreement
with the result of \cite{Inoue:jq}.

The only master integral of the transcendentality four contributing
to the result is:
\begin{eqnarray}
&&H_4=\frac{m_t^4}{(i \pi^2)^2}\int \int d^4k_1 d^4k_2
          P_{k_1+q_1,m_t}P_{k_1+q_2,m_t}
          P_{k_2+q_1,m_t}P_{k_2+q_1,m_t}
          P_{k_1-k_2,m_t} P_{k_2,m_t}
   \nonumber \\
&& \nonumber \\
&&~~= \frac{2\theta_t^2}{(1-{\theta}_t)^3(1+{\theta}_t)}
 \left[
 \left( \frac72 \Li2({\theta}_t)+2 \Li2(-{\theta}_t)+
          \zeta_2 \right) \ln^2 {\theta}_t
 \right.
 \nonumber \\
&& \nonumber \\
&&~~\left.-4\left(\Li3({\theta}_t^2)-\zeta_3 \right) \ln {\theta}_t
  + \frac{1}{48} \ln^4{\theta}_t  +\frac92 \zeta_4
   +27 \Li4({\theta}_t)+36 \Li4(-{\theta}_t) 
    \right ],
\label{H4}
\end{eqnarray}
where $P^{-1}_{k,m} = k^2-m^2+i \epsilon$ and $q_1^2=q_2^2=0$,
$(q_1-q_2)^2=M_H^2$.
For the calculation of this result a method to some extent resembling the
`Risch algorithm' \cite{Risch1} was developed.
To conclude, we stress that the availability of analytic results is
of great value since they allow to cover the whole
kinematical domain with one expression, serve as checks for approximation,
allow to represent the analytic properties of the amplitudes and can
also be used for numerical evaluation in all domains of interest.

\vspace{1.cm}

{\bf Acknowledgment}
J.F. wants to thank DESY Zeuthen for various invitations.

\end{document}